\documentclass[12pt]{article}
\usepackage{amssymb}
\begin{document}
\title{On renormalization of Poisson--Lie T--plural sigma models}
\author{Ladislav Hlavat\'y, Josef Navr\'atil, Libor \v Snobl }

\def\tuc{\bf}
\def\ba{\begin{eqnarray}}
\def\ea{\end{eqnarray}}
\def\be{\begin{equation}}
\def\ee{\end{equation}}
\def\lbl{\label}
\def \rf  {(\ref}

\def\eqn{equation}
\def\cond{condition}
\def\tfn{transformation}
\def\soln{solution}
\def\fn{function}
\def\sm{$\sigma$--model}
\def\pl{Poisson--Lie}
\def\pltfn{Poisson--Lie transformation}
\def\dd{Drinfel'd double}
\def\rgp{renormalization group}
\def\rgn{renormalization group equation}
\def\-1{^{-1}}
\def\half{\frac{1}{2}}
\def\coor{coordinate}
\def\real{{\mathbb{R}}}
\def\compl{{\mathbb{C}}}
\def\unit{{\bf 1}}

\def\e{{\rm e}}
\def\cd{{\mathcal D}}
\def\cg{{\mathcal G}}
\def\tcg{\widetilde{{\mathcal G}}}
\def\tx{\widetilde{X}}
\def\ce{{\cal {E}}}

\def\el{{e}^{L}}
\def\vr{{v}^{R}}
\def\vl{{v}^{L}}
\def\dpm{{\partial_\pm}}
\def\rd{{\rm d}}
\def\ttil{\tilde{T}}
\def\that{\widehat{T}}
\def\ghat{\hat{g}}
\def\htil{\tilde{h}}

\def\text{}
\def\wh{\hat}
\def\wt{\widetilde}
\def\sm{$\sigma$--model}
\def\pltp{Poisson--Lie T--pluralit}
\def\pltd{Poisson--Lie T--dualit}
\def\dd{Drinfel'd double}
\def\mt{Manin triple}
\maketitle {\begin{center}\noindent Czech Technical University in Prague, Faculty of Nuclear Sciences and \\ Physical
Engineering,  B\v rehov\'a 7, 115 19 Prague 1, Czech Republic\end{center}}

\begin{abstract} {Covariance of the one-loop renormalization group equations with respect to Poisson--Lie T--plurality of sigma models is discussed. The role of ambiguities in renormalization group equations of Poisson--Lie sigma models with truncated matrices of parameters is investigated.}
\end{abstract}

\section{Introduction}
One--loop renormalizability of \pl{} dualizable \sm s and their \rgn s  were derived in \cite{valklisque}.  Covariance of the \rgn s with respect to \pltd y was proven in \cite{sfesia}. That suggests that also properties of quantum \sm s can be given in terms of \dd s and not their decompositions into Manin triples. This was indeed claimed in~\cite{sfesiathom} where a renormalization on the level of sigma models defined on Drinfel'd double was proposed. A natural way to independently verify this claim would be to extend the proof of covariance of~\cite{sfesia} to \pltp y.

Unfortunately, transformation properties of the structure constants and the matrix $M$ (parameters of the models) under the \pltp y are much more complicated than in the case of T--duality. That's why we decided to check it first on examples using our lists of 4-- and 6--dimensional Drinfel'd doubles and their decompositions into Manin triples \cite{hlasno:2dim,snohla:ddoubles}.

It turned out that the renormalization group equations of \cite{valklisque,sfesia} are indeed invariant under Poisson--Lie T--plurality. The equivalence of the renormalization flows of the models on the Poisson--Lie group of~\cite{sfesia} and on the Drinfel'd double~\cite{sfesiathom} also holds in all cases studied so far provided one is careful in the interpretation of the formulas in different parts of~\cite{sfesiathom}, see Section~\ref{siampos}.

An assumption in the renormalizability proof \cite{valklisque} is that there
is no a priori restriction on elements of matrix $M$ that together with the structure of
the \mt{} determine the models.  It was noted in \cite{sfesia,sfe:dic2dim} that the \rgn s
need not be consistent with truncation of the parameter space. On the other hand there is
some freedom in the \rgn s and we are going to show how they can be used in the choice of
one--loop $\beta$ functions for a given truncation.

\section{Review of \pltp y}\label{review}

For simplicity we shall consider \sm s without spectator fields, i.e. with target manifold isomorphic to a group. Let $G$ be a Lie group and ${\cal{G}}$ its Lie algebra. Sigma model on the group $G$ is given by the classical action
    \be\label{SEg}
        S_{E}[g]=\int d^{2}x R_{-}(g)^{a}E_{ab}(g)R_{+}(g)^{b},
    \ee
where $g:\mathbf{R}^{2}\rightarrow G,(\sigma_+,\sigma_-)\mapsto g(\sigma_+,\sigma_-)$,
$R_{\pm}$ are right-invariant fields $R_{\pm}(g):=(\partial_{\pm}gg^{-1})^{a}T_{a}\in
{\cal{G}}$ and $E(g)$ is a certain bilinear form on the Lie algebra $\cg$, to be specified
below.

The \sm s that can be transformed by the \pltd y are formulated (see
\cite{klse:dna,kli:pltd}) by virtue of \dd{} $D\equiv(G|\tilde G)$
-- a Lie group whose Lie algebra $\cd$ admits a decomposition
$\cd=\cg\dotplus\tcg$ into a pair of subalgebras maximally isotropic
with respect to a symmetric ad-invariant nondegenerate bilinear form
$\langle\, .\,,.\,\rangle $. These decompositions are called Manin
triples.

The matrices $E(g)$ for such \sm s are of the form \be E(g)=(M+\Pi(g))^{-1}, \ \ \
\Pi(g)=b(g)\cdot a\-1(g) = -\Pi(g)^t,\label{Fg}\ee where $M$ is a constant matrix and
$a(g),b(g)$ are submatrices of the adjoint representation of the subgroup $G$ on the Lie
algebra $\cd$ defined as \be\label{adgt} gTg\-1\equiv Ad(g)\triangleright T=a\-1(g)\cdot
T,\ \ \ \ g\tilde Tg\-1\equiv Ad(g)\triangleright \tilde T =b^t(g)\cdot T+ a^t(g)\cdot
\tilde T, \ee where $T_a$ and $\ttil^a$ are elements of dual bases of $\cg$ and $\tcg$,
i.e. $$\langle\,T_a ,\,T_b\,\rangle=0,\ \ \langle\,\ttil^a ,\,\ttil^b\,\rangle=0,\ \
\langle\,T_a ,\,\ttil^b\,\rangle=\delta_a^b .$$

Origin of the \pltp y \cite{klse:dna,unge:pltp} lies in the fact that
in general several decompositions (Manin triples) of the \dd{} may
exist. Let $\cd=\hat\cg\dotplus\bar\cg$ be another decomposition of
the Lie algebra $\cd$ into maximal isotropic subalgebras. The dual
bases of $\cg,\tcg$ and $\hat\cg,\bar\cg$ are related by the linear
\tfn\begin{equation}\label{pqrs}
    \left(\matrix{T \cr\tilde T\cr} \right)= \left(\matrix{K&Q \cr W&S \cr} \right) \left(\matrix{\hat T\cr
\bar T\cr} \right), \end{equation} where the matrices $K,\, Q,\, W,\, S$ are chosen in such a way that the structure of the Lie algebra $\cd$ in the basis $(T_a,\ttil^b)$
\begin{eqnarray}\label{liestruc}  [T_a,T_b] &=& {f_{ab}}^c T_c,\nonumber\\ {}
[\tilde T^a,\tilde T^b] &=& {{\tilde f}^{ab}}{}_c\tilde T^c,\\ {}
[\tilde T^a,T_b] &=&   {f_{bc}}^a \tilde T^c - {{\tilde f}^{ac}}{}_b
T_c\nonumber
\end{eqnarray}transforms to a similar one where $T\rightarrow\that,\ \ttil\rightarrow\bar T$ and
the structure constants $f,\, \tilde f$ of $\cg$ and $\tcg$ are replaced by the structure
constants $\hat f,\,\bar f$ of $\hat \cg$ and $\bar \cg$. The duality of both bases
requires \be\label{trinv} \left(\matrix{K&Q \cr W&S \cr} \right)\-1=\left(\matrix{S^t&Q^t
\cr W^t&K^t \cr} \right).\ee The \sm {} obtained by the \pltp y is defined analogously to
(\ref{SEg})-(\ref{Fg}) where
\[ \wh E(\ghat)=(\wh M+\wh\Pi(\ghat))^{-1}, \ \ \
\wh\Pi(\ghat)=\wh b(\ghat)\cdot \wh a\-1(\ghat) = -\wh\Pi(\ghat)^t,
\]
\begin{equation}\label{E0hat}
 \hat M=(M\cdot Q+S)\-1\cdot (M\cdot K+W) =(K^t\cdot M-W^t)\cdot (S^t-Q^t\cdot M)\-1.
\end{equation}
 The transformation (\ref{E0hat}) $M\mapsto\hat M$ is obtained when the subspaces
 ${\cal E}^\pm={ span}\{E^\pm_a\}_{a=1}^n$ spanned
by \begin{equation}\label{subspsE}
    E_a^+:=T_a+M^{-1}_{ab}\tilde T^b,\ \     E_a^-:=T_a-M^{-1}_{ba}\tilde T^b
\end{equation}
are expressed as ${\cal E}^+={ span}\{\hat T_a+\hat M^{-1}_{ab}\bar T^b\}_{a=1}^n$, ${\cal
E}^-={ span}\{\hat T_a-\hat M^{-1}_{ba}\bar T^b\}_{a=1}^n$.

Classical solutions of the two \sm s are related by two possible decompositions of $l\in
D$,
\begin{equation}\label{lgh}
    l=g\htil =\ghat\bar h .
\end{equation}
Examples of explicit solutions of the \sm s related by the \pltp y were given in
\cite{hlahytur}. The \pltd y is a special case of \pltp y with $K=S=0,\ Q=W=\unit$.

It is useful to recall that several other conventions are used in the literature. {E.g.,}
the action in \cite{sfesia,sfesiathom} is defined as
 \be\label{SSfetsos}
        S[g]=\int d^{2}x L_{+}(g)\cdot(M+\bar\Pi(g))\-1\cdot L_{-}(g)^{b},
    \ee
where $\bar\Pi(g)=b^t(g)\cdot a(g)=\Pi(g\-1)$. The {transition} between the actions
(\ref{SEg}) and (\ref{SSfetsos}) is given by $g\leftrightarrow g\-1,\ M\leftrightarrow
M^t.$\medskip

The one--loop \rgn s for \pl{} dualizable \sm s were found in \cite{valklisque}. In our
{notation} it reads
\begin{equation}\label{rgeqn1}
    \frac{dM^{ba}}{dt}=r^{ab}(M^t).
\end{equation}
Notice that the equation (\ref{rgeqn1}) appears in \cite{valklisque,sfesia} without
transposition of $M$ on both sides of the equation due to different formulations of the \sm{} action (\ref{SEg}) vs. (\ref{SSfetsos}).

The matrix valued function $r^{ab}$ is defined as
\begin{equation}r^{ab}(M)={R^{as}}_t(M){L^{tb}}_s(M)\end{equation}\begin{equation}\label{RR}
    {R^{ab}}_c(M)=\half(M_S\-1)_{cd}\left( {A^{ab}}_e M^{de}+{B^{ad}}_e M^{eb}-{B^{db}}_e M^{ae}\right),
\end{equation}
\begin{equation}\label{LL}
    {L^{ab}}_c(M)=\half(M_S\-1)_{cd}\left( {B^{ab}}_e M^{ed}+{A^{db}}_e M^{ae}-{A^{ad}}_e M^{eb}\right),
\end{equation}
\begin{equation}\label{AB}
    {A^{ab}}_c={{\tilde f}^{ab}}{}_c-{f_{cd}}^aM^{db},\ \ \ \  {B^{ab}}_c={{\tilde f}^{ab}}{}_c+M^{ad}{f_{dc}}^b,
\end{equation}
\begin{equation}\label{MinvE0}
   M_S=\half(M+M^t).
\end{equation}
It was shown in \cite{sfesia} that the \eqn~(\ref{rgeqn1}) is covariant with respect to the
\pltd y, i.e., it is equivalent to
\begin{equation}\label{rgeqn2}
    \frac{d\tilde M^{ba}}{dt}=\tilde r^{ab}(\tilde M^t)
\end{equation}
obtained by
\begin{equation}\label{pltdy} f\rightarrow\tilde f,\ \tilde f\rightarrow f,\
M\rightarrow \tilde M=M\-1.
\end{equation}
One expects that the \eqn s (\ref{rgeqn1}) are covariant also with respect to
the \pltp y when
\begin{equation}\label{pltpy}    f\rightarrow\hat f,\ \tilde f\rightarrow\bar f,\
M\rightarrow \wh M,
\end{equation}
where the transformation of $\wh M$ under plurality is given by (\ref{E0hat}). We have checked the invariance on numerous examples of Poisson--Lie T--plurality using 4-- and 6--dimensional Drinfel'd doubles and their decompositions into Manin triples of \cite{hlasno:2dim,snohla:ddoubles} and have found no counterexamples.

\section{Relation to the \rgn s on the \dd}\label{siampos}
The above presented renormalization equation (\ref{rgeqn1}) shall be compared to the \rgn s derived
in \cite{sfesiathom} on the whole \dd{}
\begin{equation}\label{328}
    \frac{dR_{AB}}{dt}=S_{AB}(R,h)=\frac{1}{4}
(R_{AC}R_{BF}-\eta_{AC}\eta_{BF}) (R^{KD}R^{HE}-\eta^{KD}\eta^{HE}) h_{KH}{}^Ch_{DE}{}^F
\end{equation} for the symmetric matrix $R$. For a given decomposition of the \dd{} into a \mt{} $(G|\tilde G)$,
the structure constants $h$ of the \dd{} are given by the structure constants $f,\tilde f$ of the subalgebras of the  \mt{} $h=h(f,\tilde f)$ as in equation~(\ref{liestruc}). The matrix $R$ is related to the matrix $M$, which defines the \sm{} on the group G, by
\begin{equation}\label{420}
   R_{AB} = \rho_{AB}(M)=\left(
           \begin{array}{cc}
             \tilde M_s - B \tilde M_s^{-1}B & -B \tilde M_s^{-1}  \\
             \tilde M_s^{-1} B & \tilde M_s^{-1} \\
           \end{array}
         \right),
\end{equation} where $$B=  \half\left[M^{-1}-(M^{-1})^t\right],\ \ \tilde M_s=\half\left[M\-1+(M\-1)^t\right],$$
$$ \ R^{AB}=(R\-1)^{AB},\ R\-1=\eta\cdot R\cdot\eta, $$and
\begin{equation}\label{eta}
 \eta_{AB} =    \langle T_A|T_B\rangle =  \left(\begin{array}{cc}
  \bf{0} & \mathbb{I}_{d_G\times d_{G}} \\
  \mathbb{I}_{d_G\times d_G} & \bf{0}
\end{array}\right),
\end{equation}where $T_A=\{T_i,\tilde T^j\}.$
It is easy to show that due to (\ref{420}) the equivalence of (\ref{328}) and
(\ref{rgeqn1}) where $r^{ab}=r^{ab}(M,f,\tilde f)$ requires
\begin{equation}\label{kompatibility rgns}
    S_{AB}\left(\rho(M),h(f,\tilde f)\right)=\frac{\partial \rho_{AB}}{\partial M^{ab}}(M)\,r^{ba}(M^t,f,\tilde
    f).
\end{equation}
Be aware of the presence of transpositions on the right--hand side.

{ By construction -- cf.~the \eqn~(4.15) of \cite{sfesiathom} -- the matrix $M$ which is put into the
equation~(\ref{420}) (and thus appears in the equation~(\ref{328})) transforms under T--plurality as in
(\ref{E0hat}), i.e. agrees with the convention used here for the sigma model of the form (\ref{SEg}).
However, the sigma models on the Poisson--Lie groups in~\cite{sfesiathom} are expressed in a different
convention, as in the equation (\ref{SSfetsos}) here. Thus, a tacit transposition of the matrix $M$ is
necessary when comparing the renormalization group flows on the double and on the individual
Poisson--Lie subgroup in~\cite{sfesiathom}. Taking this fact into consideration we were able to recover
the examples presented in~\cite{sfesiathom} and also confirm the conjectured equivalence of the
renormalization group equations~(\ref{328}) and~(\ref{rgeqn1}) in all the investigated 4-- and
6--dimensional \dd s. }

\section{Non--uniqueness of the \rgn s}It was noted in the paper \cite{valklisque} that there is
a certain ambiguity in the one--loop \rgn s. Namely, the flow given by the
\eqn~(\ref{rgeqn1}) is physically equivalent to the one given by the \eqn
\begin{equation}\label{rgeqn}
    \frac{dM^{ba}}{dt}=r^{ab}(M^t)+{R^{ab}}_c(M^t)\,\xi^c,
\end{equation}
where  $\xi^c$ are arbitrary functions of the renormalization scale $t$.

The origin of this arbitrariness in  $\xi^c$ lies in the fact that the metric and B--field
are determined up to the choice of coordinates, i.e. up to a diffeomorphism, of the group
$G$ viewed as a manifold. In our case we may in addition require that the transformed
action takes again the form~(\ref{SEg})-(\ref{Fg}) for some matrix $M'$. On the other hand,
we do not have to require the diffeomorphism to be a group homomorphism because the group
structure plays only an auxiliary role in the physical interpretation.

For example, in the particular case of semi--Abelian double, i.e. {$\tilde f=0$,} $\Pi=0$,
with a symmetric matrix $M$, the left translation by an arbitrary group element
$h=\exp(X)\in G$, i.e. replacement of $g$ by $hg$ in the action~(\ref{SEg}), leads to the
new matrix $M'=Ad(h) \cdot M \cdot Ad(h)$, specifying a metric physically equivalent to the
original one. Such a diffeomorphism is generated by the flow of the left--invariant vector
field $X$. For general Manin triples and matrices $M$ similar transformations are generated
by more complicated vector fields parameterized by $\xi^c$, as was found in
\cite{valklisque}. Thus the renormalization group flows~(\ref{rgeqn}) differing by the
choice of $\xi^c$ are physically equivalent. Consistency under the \pltp y requires that
the functions $\hat \xi^c$ for the plural model satisfy
\begin{equation}\label{tfnxibar}
   \wh R(\hat M^t)\cdot\hat\xi=  ( S-M^t\cdot Q)\-1 \cdot (R(M^t)\cdot\xi)\cdot(K+Q\cdot \wh M^t).
\end{equation}
For the \pltd y this formula simplifies to $$\tilde R( \tilde M^t)\cdot(\tilde\xi+\tilde
M^t\cdot\xi)=0.$$

The {freedom} in the choice of the functions $\xi^a$ can be employed when compatibility of
the \rgn {} flow with a chosen ansatz (truncation) for the matrix $M$ is sought.

\subsection{Renormalizable \sm s for $M$ proportional to the unit or diagonal matrix} The
simplest ansatz for the constant matrix is $M=m\unit$ where $\unit$ is the identity matrix
and  $m\neq 0$. As mentioned in the Introduction, truncation or symmetry of the constant
matrix $M$ that determines the background of the \sm{} often contradicts the form of the
r.h.s of the \rgn s (\ref{rgeqn1}). On the other hand, the freedom in the choice of $\xi^c$
in (\ref{rgeqn}) may help to restore the renormalizability. It is therefore of interest to
find consistency conditions for the \rgn s for the \sm s given by this simple $M$.

Two--dimensional \pl{} \sm s  are given by \mt s generated by Abelian or solvable Lie
algebras with Lie products
\begin{equation}\label{bianchimt2a}
   [T_1,T_2]=a\,T_2,\ [\ttil^1,\ttil^2]=\tilde a\,\ttil^2,\ a\in\{0,1\},\ \tilde a\in\real
\end{equation}   or
\begin{equation}\label{bianchimt2b}
[T_1,T_2]=T_2,\ [\ttil^1,\ttil^2]=\ttil^1\end{equation}

In the former case, the \eqn{}  (\ref{rgeqn}) for $M=m\unit$ reads
\begin{equation}\label{rgeqn2a}
\left(
\begin{array}{cc}
 \frac{dm}{dt} & 0 \\
 0 & \frac{dm}{dt}
\end{array}
\right)=\left(
\begin{array}{cc}
 a^2m^2 -\tilde a^2& (a\, m+\tilde a) \xi^2
   \\
 0 & -(a\, m+\tilde a)  \xi^1
\end{array}
\right)
\end{equation}
so that we generically get $\xi^1=\tilde a- a\, m,\ \xi^2=0$ and the \rgn{} is
${dm}/{dt}=a^2m^2-\tilde a^2 $. In the special case $a=1, \, m=-\tilde a$ the r.h.s. of the
equation~(\ref{rgeqn2a}) vanishes for all choices of $\xi^k$, i.e. there is no renormalization. Notice
that had we allowed a diagonal ansatz
\begin{equation}\label{diagonalM}
M=\left(\begin{array}{cc} m_1 & 0 \\ 0 & m_2 \end{array} \right)
\end{equation}
instead of the multiple of the unit matrix, the restriction on the value of $\xi^1$ would disappear
and the \rgn{} would take the form
\begin{equation}
 \frac{dm_1}{dt} = -\tilde a^2+m_1^2 a^2, \qquad \frac{dm_2}{dt} = -\frac{m_2}{m_1}\xi^1 (\tilde a+m_1 a).
\end{equation}

For the Manin triple~(\ref{bianchimt2b}), the \eqn{} (\ref{rgeqn}) reads
\begin{equation}\label{rgeqn2b}
\left(
\begin{array}{cc}
 \frac{dm}{dt} & 0 \\
 0 & \frac{dm}{dt}
\end{array}
\right)=\left(
\begin{array}{cc}
 m^2+\xi^2& m\,(\xi^2-1)
   \\
 m- \xi^1 & -1- m\, \xi^1
\end{array}
\right)
\end{equation}
and no choice of $\xi^1, \xi^2$ satisfies the \eqn{} (\ref{rgeqn2b}). Therefore the \pl{} \sm {} given by \mt{}
(\ref{bianchimt2b}) is not renormalizable with $M$ kept proportional to the unit matrix. The situation changes when we allow general diagonal form (\ref{diagonalM}) of the matrix $M$. Then the \rgn{} becomes
\begin{equation}\label{rgeqn2bII}
\left(
\begin{array}{cc}
 \frac{dm_1}{dt} & 0 \\
 0 & \frac{dm_2}{dt}
\end{array}
\right)=\left(
\begin{array}{cc}
 m_1^2+\frac{m_1}{m_2}\xi^2& m_1\,(\xi^2-1)
   \\
 m_1- \xi^1 & -1- m_2\, \xi^1
\end{array}
\right)
\end{equation}
which allows the flow
$$\frac{dm_1}{dt} = m_1^2+\frac{m_1}{m_2}, \qquad \frac{dm_2}{dt} = -1-m_1 m_2$$
respecting the diagonal ansatz (\ref{diagonalM}) for the unique choice  $\xi^1 = m_1$,
$\xi^2 = 1$.

Consistency of the one--loop \rgn s for three--dimensional \pl{} \sm s with $M$
proportional to the unit matrix fixes $\xi^3=0$ and is consistent with the choice $\xi^2=0$
(unique in some cases). It exists for the following \mt s and choices of $\xi^1$ and/or $m$
\begin{eqnarray}
\label{beglist}  (1|1):\ \ \frac{dm}{dt} &=& 0,\ \ \ \xi^1=0,  \\
(3|3.i|b):\ \ \frac{dm}{dt} &=& 0,\ \ \ \xi^1=0,\  m=\pm b,  \\
  (5|1):\ \ \frac{dm}{dt} &=& 2m^2,\ \ \ \xi^1=2m,  \\
(6_0|5.iii|b):\ \ \frac{dm}{dt} &=& 0,\ \ \ \xi_1=0,\  m=\pm b,  \\
(6_a|6_{1/a}.i|b):\ \ \frac{dm}{dt} &=& 0,\ \ \ \xi_1=0,\  m=\pm b/a,  \\
(6_a|6_{1/a}.i|b):\ \ \frac{dm}{dt} &=& 2b^2(a^2-\frac{1}{a^2}),\ \ \xi_1=-2b(a+\frac{1}{a}),\  m=- b,  \\
(7_a|1):\ \ \frac{dm}{dt} &=& 2a^2m^2,\ \ \ \xi^1=2am,\ \  a\geq 0 \\
(7_a|7_{1/a}|b):\ \ \frac{dm}{dt} &=& 2(m^2-b^2),\ \ \ \xi^1=2(m-b),\ \ a=1,  \\
  (9|1):\ \ \frac{dm}{dt} &=& -m^2/2,\ \ \ \xi^1=0,  \\
 \label{endlist} (9|5|b): \frac{dm}{dt}&=&-\half m^2-2 b^2,\ \xi^1=-2b
\end{eqnarray}
and their duals (for notation of $(X|Y)$  or $(X|Y|b)$ see \cite{snohla:ddoubles}).
Renormalization of \pl{} \sm s given by other six--dimensional \mt s is not consistent with
the assumption $M$ proportional to identity, i.e. renormalization spoils the ansatz.

We have also investigated three--dimensional \sm s with general diagonal matrices $M$ but the list of
renormalizable models is rather long so that we do not display it here.

We notice that the list of renormalizable three--dimensional \pl{} \sm s with $M$ proportional to the
unit matrix is in agreement with the results obtained in \cite{hlasno:3dsm1}. There the conformally
invariant \pl{} \sm s, i.e. those with vanishing $\beta$--function, were studied and the sigma models
with diagonal $M$ and constant dilaton field  were obtained. They appear in the above constructed list
with vanishing r.h.s of the \rgn{}.

\section{Conclusions}
We have discussed the transformation properties of the renormalization group flow under Poisson--Lie T--plurality.

Originally we expected on the basis of our previous experience with the Poisson--Lie T--duality and
T--plurality that it should possible to generalize the proof of the equivalence of the renormalization
group flows (\ref{rgeqn1}) of Poisson--Lie T--dual sigma models \cite{sfesia} to the case of
Poisson--Lie T--plurality. { Unfortunately, this task proved to be beyond our present means due the
relative complexity of the transformation formula~(\ref{E0hat}) compared to the duality
case~(\ref{pltdy}). Thus, we resorted to investigation of the invariance properties of the
renormalization group flows on low--dimensional examples. We have found no contradiction with the
hypothesis that the renormalization group flows as formulated in \cite{sfesia} are equivalent under the
\pltp y and with the claim that the renormalization renormalization flows of the models on the
Poisson--Lie group and on the Drinfel'd double are compatible.}

Next, we studied whether the freedom in the choice of functions $\xi^c$ in the \rgn s
(\ref{rgeqn}) can be employed to preserve chosen ansatz of the matrix $M$ during the the
renormalization group flows. It turned out that indeed this ambiguity often enables to stay
within the diagonal ansatz for the matrix $M$.

\section*{Acknowledgement}
This work was supported by the grant  RVO68407700 and the research plan MSM6840770039 of
the Ministry of Education of the Czech Republic (L.H. and L.\v S) and by the Grant Agency
of the Czech Technical University in Prague, grant No. SGS10/295/OHK4/3T/14 (J.N.).

We are grateful to Konstadinos Sfetsos and Konstadinos Siampos for e-mail discussions that helped to
pinpoint the differences in notation and corresponding reformulations of the \rgn s.

\end{document}